\documentclass[nohyper,12pt,letterpaper]{JHEP3}
\usepackage{epsfig}
\title{Massive IIA flux compactifications and U-dualities}
\author{T.\,Banks\\ Department of Physics and SCIPP\\
University of California, Santa Cruz, CA 95064\\
E-mail: \email{banks@scipp.ucsc.edu}\\
{\it and}\\
Department of Physics and NHETC, Rutgers University\\
Piscataway, NJ 08540}
\author{K.\,van den Broek\\
Department of Physics and NHETC, Rutgers University\\
Piscataway, NJ 08540\\
E-mail: \email{korneel@physics.rutgers.edu}} \abstract{We attempt to
find a rigorous formulation for the massive type IIA orientifold
compactifications of string theory introduced in \cite{DeWolfe}.  An
approximate double T-duality converts this background into IIA
string theory on a twisted torus, but various arguments indicate
that the back reaction of the orientifold on this geometry is large.
In particular, an AdS calculation of the entropy suggests a scaling
appropriate for $N$ M2-branes, in a certain limit of the
compactification, though not the one studied in \cite{DeWolfe}. The
M-theory lift of this specific regime is not 4 dimensional. We
suggest that the generic limit of the background corresponds to a
situation analogous to F-theory, where the string coupling is small
in some regions of a compact geometry, and large in others, so that
neither a long wavelength 11D SUGRA expansion, nor a world sheet
expansion exists for these compactifications. We end with a
speculation on the nature of the generic compactification.}

\received{} \accepted{}

\preprint{
hep-th/0611185\\
\\
RUNHETC-06-29, SCIPP-06-12 \\}

\begin{document}
%%%%%%%%%%%%%%%%%%%%%%%%%%%%%%%%%%%%%%%%%%%%%%%%%%%%%%%%%%%%%%%%%%%%%%%%%%%%
% Table of contents automatic !!!                                          %
%%%%%%%%%%%%%%%%%%%%%%%%%%%%%%%%%%%%%%%%%%%%%%%%%%%%%%%%%%%%%%%%%%%%%%%%%%%%

%=====================
%=====================
\section{Introduction}
Flux compactifications \cite{FluxCompactifications} provide the
arena for most of the discussions of the String Landscape as well as
modern approaches to string phenomenology.   The discussion of these
compactifications is generally carried out in low energy effective
field theory \cite{GrimmLouis} \cite{Tom}, despite the fact that
they all contain orientifold singularities. Further, there is no
perturbative world sheet treatment of these backgrounds. Recently,
DeWolfe {\it et al.}\cite{DeWolfe} introduced a sequence of models
characterized by an integer $N$. Earlier work on similar type IIA
flux compactifications was done in \cite{VZetal}. The DeWolfe {\it et al.}
compactifications are classical solutions of Type IIA SUGRA, with a
singular orientifold source and a variety of Ramond-Ramond and Neveu
Schwarz fluxes.  The parameter $N$ is related to the value of
certain quantized fluxes, and may be taken arbitrarily large.  This
is in marked contrast to typical flux compactifications, where
fluxes are bounded \cite{FiniteVacua}. The authors of \cite{DeWolfe}
argue that for large $N$ the moduli can be stabilized at values
where all radii are large compared to the string scale, and the
string coupling is small. Furthermore, the four non-compact
directions are an AdS space with a radius $R_{AdS}$ whose ratio to
the compactification scale grows with $N$. The latter property is in
marked contrast to the sequences of models treated in the AdS/CFT
correspondence.

Our aim in this paper is to investigate further the models of
\cite{DeWolfe}, and to determine whether they admit a systematic low
energy field theory expansion (see also \cite{AcharyaEtAl}) and/or a
weakly coupled string expansion.   The inevitable orientifold of
flux compactifications is one potential barrier to an effective
field theory treatment\footnote{G.~Moore and S.~Ramanujam have
emphasized to us the problems with the back reaction of the
orientifold, which they have analyzed extensively in the context of
the original DeWolfe {\it et al.} solutions\cite{Sridhar}.}.   In
addition, these models contain a ten form flux $F_0$, and correspond
to solutions of the massive Type IIA SUGRA Lagrangian.   It is well
known that quantization of $F_0$ is a problem for effective field
theory, and that the massive Type IIA string theory does not have a
perturbative world sheet expansion (the D8-brane solution of this
theory has a string coupling which grows at infinity).   In
addition, the effective field theory treatment has the usual problem
of orientifold singularities.   Thus despite apparently small
parameters, it is far from clear that there is a systematic large
$N$ expansion of this system.

We approach this problem indirectly.   Ignoring the back reaction of
the orientifold, we perform a double T-duality on the DeWolfe {\it et al.}
background\footnote{We work in the orbifold limit.   The authors of
\cite{DeWolfe} took pains to show that the blow up moduli of the
orbifold can be stabilized at large values of the radii of shrinking
cycles. We address the analogous question in the T-dual picture.}.
The result is Type IIA string theory on a twisted torus, with flux
only in the four AdS directions. Despite the fact that this
configuration does not satisfy tadpole cancellation, the T-duality
is a legitimate operation on the orbifold CFT.   We then restore
tadpole cancellation in the T-dual picture (the formal dual of
the original orientifold).

\paragraph{Regime with one large 4-flux: \\}
We argue that the resulting model in this regime does not have a weakly
coupled Type IIA world sheet expansion. In this limit, from the
point of view of DeWolfe {\it et al.}, the string coupling remains
weak, the scales of both $AdS_4$ and the compact manifold are large,
and the Kaluza-Klein radius is parametrically smaller.   However,
some cycles on the compact manifold shrink to zero size, and this is
not a limit in which DeWolfe {\it et al.} would claim to have a
controlled expansion.  In the limit where we only turn on one four
form flux, the fixed temperature entropy computed from the AdS
geometry scales like $N^{3/2}$ as one would expect from a large
number of M2-branes. We show that this is explained in the T-dual
IIA picture by a large number of D2-branes sitting at the
orientifold locus, where the string coupling is large. The D2/M2
world volumes are in the AdS directions.   In typical orientifold
compactifications that have been studied in string theory, those
with a known world sheet expansion, the effect of the orientifold is
confined to a region of order string scale.   Here we argue that
this is not the case, since the parameter $N$, which apparently
tunes the string coupling to be small, in fact counts a large number
of branes near the orientifold singularity.   We argue that in fact
the strong coupling region completely dominates the geometry in the
single flux limit. The resulting theory for large $N$ is M-theory on
$AdS_4 \times M_7$, where $M_7$ is a manifold of weak $G_2$
holonomy.   The AdS and compact radii scale the same way with $N$.

\paragraph{Regime with all 4-flux large: \\}
For the generic regime of the background, we find that 11D SUGRA is not a valid approximation. This
is a consequence of the small string coupling found by DeWolfe et
al., combined with the observation that the AdS radius is much
larger than that of the compact manifold computed using our naive
T-duality rules. Thus, in this region where DeWolfe {\it et al.}
claimed a systematic expansion, many features of their picture are
valid. However, our picture also includes large numbers of D-branes
sitting at the orientifold locus in the regime where all fluxes are
large. We argue that the weak coupling approximation breaks down in
a vicinity of the orientifold whose size scales like $N^{1/20} l_s$.
This rules out a uniform weak coupling expansion in the large N
limit.   Furthermore, if we apply 11D SUGRA to the region around the
orientifold, it suggests that this region actually blows up to a
seven manifold whose radius of curvature is of order the AdS radius.
In the conclusions, we also provide a heuristic explanation of the
peculiar $N^{9/2}$ entropy scaling of the regime with all fluxes
large.   This argument also seems to require a compact manifold with
volume much larger than that suggested by De Wolfe {\it et al.}.

Our conclusion is that the generic DeWolfe {\it et al.} configuration
probably exists as a valid model of quantum gravity in $AdS_4$.
However, it is unclear to us whether is has a compactification
radius parametrically smaller than the AdS radius. No existing
approximation scheme computes its large $N$ expansion. Different
approximations, apparently valid in different regions of the compact
manifold suggest different values for the ratio of scales.   The
problem of different approximation schemes for different regions is
somewhat analogous to F-theory solutions for fluxless
compactifications.  However, the large supersymmetry algebra of
F-theory compactifications provides reliable computational tools,
which are absent for these models.

The paper is organized as follows. In section 2, we review the
DeWolfe {\it et al.} background. In section 3, we transform the background
by a double T-duality, using the approximations noted above. This
allows us to eliminate the massive type IIA flux. We also comment on
the approximate character of the transformation. Section 4 deals
with the Bianchi condition for the dualized background. In section
5, we will argue that the DeWolfe {\it et al.} solution with one large flux should be
considered in an M-theory setting. We will explicitly lift the
dualized background to M-theory. Section 6 will detail some of the
aspects of the obtained 11D SUGRA solution. We will discuss its
interpretation as a stack of M2-branes.  We conclude in section 7
where we speculate on the nature of the generic DeWolfe {\it et al.} compactification.
Appendix \ref{TdualDictionary} and \ref{TdualComputation} give some
more details on the double T-duality transformation of the DeWolfe
{\it et al.} background, while Appendix \ref{IIA-MthDictionary} reviews
the formulas to lift the background to M-theory.

%======================================
%======================================
\section{The DeWolfe {\it et al.} background}

%----------------------------------------------------------------------------------------------
\subsection{The metric, fluxes and discrete symmetries of the solution} \label{DeWolfeSolution}
In \cite{DeWolfe}, DeWolfe {\it et al.} describe an infinite set of
$\mathcal{N} = 1$ solutions of massive type IIA SUGRA \cite{Romans}.
The compact manifold in their solution is $T^2 \times T^2 \times
T^2$, modded out by three discrete symmetries:
\begin{itemize}
  \item $\Omega_p (-1)^{F_L} \sigma$ with $\sigma: z_i \rightarrow -\bar{z}_i$
  \item $T: (z_1,z_2,z_3) \rightarrow (\alpha^2 z_1, \alpha^2 z_2, \alpha^2 z_3)$
  \item $Q: (z_1,z_2,z_3) \rightarrow (\alpha^2 z_1 + \frac{1+\alpha}{3},
                                       \alpha^4 z_2 + \frac{1+\alpha}{3},
                                                z_3 + \frac{1+\alpha}{3})$
\end{itemize}
with $\alpha = e^{2\pi i/6}$. The resulting space is orientifolded
$T^6 / \mathbb{Z}_3^2$. The combination of the imposed discrete
symmetries and fluxes turned on leads to a background where all
moduli are fixed. The metric and the fluxes of the background are
given by,
\begin{eqnarray}
ds^2    &=&   \gamma_1 (dx_1^2 + dx_2^2)
            + \gamma_2 (dx_3^2 + dx_4^2)
            + \gamma_3 (dx_5^2 + dx_6^2)
            + ds_{\mathrm{AdS}_4}^2                                       \label{DeWolfeSol1}  \\
H_3     &=& - 4\pi^2\alpha' h_3 \, \beta_0                                \label{DeWolfeSol1H3} \\
        &=& - 4\pi^2\alpha' h_3 \sqrt[4]{3} \sqrt{2} \,
            \bigl(
                  dx_1 \wedge dx_3 \wedge dx_5
              -   dx_1 \wedge dx_4 \wedge dx_6                                                 \nonumber \\
        & &   \qquad \qquad \qquad \:\: -\, dx_2 \wedge dx_3 \wedge dx_6
              -   dx_2 \wedge dx_4 \wedge dx_5
            \bigr)                                                                             \\
e^{\varphi}
        &=& \frac{1}{4} |h_3| \sqrt[4]{\frac{3^3 5}{|f_0 f_4^1 f_4^2 f_4^3|}}                  \\
F_4     &=& (2\pi\sqrt{\alpha'})^3   \sqrt[3]{\kappa} f_4^i \tilde{w}^i                        \\
        &=& 4 (2\pi\sqrt{\alpha'})^3 \sqrt[3]{3}
            \Bigl(
                          f_4^1 \,dx_3 \wedge dx_4 \wedge dx_5 \wedge dx_6                     \nonumber \\
        & & \qquad \qquad \quad \;\: +\, f_4^2 \,dx_5 \wedge dx_6 \wedge dx_1 \wedge dx_2      \\
        & & \qquad \qquad \quad \;\: +\, f_4^3 \,dx_1 \wedge dx_2 \wedge dx_3 \wedge dx_4
            \Bigr)                                                                             \nonumber \\
F_2     &\approx& 0                                                     \label{F2zero}         \\
F_0     &=& \frac{f_0}{2\pi\sqrt{\alpha'}}                              \label{DeWolfeSol2}    \,,
\end{eqnarray}
where $f_0, h_3, f_4^1, f_4^2, f_4^3 \in \mathbb{Z}$; $z_1 = x_1 + i x_2, \ldots$ and
\begin{equation}  \label{gamma} % work from around 03/07/06
  \gamma_i = 4\pi^2\alpha' \frac{2}{\sqrt[3]{3}}
             \sqrt{\frac{5 |f_4^1 f_4^2 f_4^3|}{|f_0|}}
             \frac{1}{|f_4^i|} \,.
\end{equation}
The tadpole cancellation condition for $F_2$ reduces to $f_0 h_3 =
-2$. Notice that we defined the R-R fluxes following the conventions
of \cite{Polchi} instead of those in \cite{DeWolfe} (see footnote 3
in \cite{DeWolfe} v3). The non-compact part of the metric,
$ds_{\mathrm{AdS}_4}^2$, is a 4-dimensional $\mathrm{AdS}_4$ space
with radius (in string frame),
\begin{equation}
  R_{\mathrm{AdS}}^2 = 4\pi^2\alpha'\, 16 \,
                       \sqrt{\frac{5^3|f_4^1 f_4^2 f_4^3|}{3^3|f_0|^3|h_3|^4}} \, .
\end{equation}
The volume of the compact manifold is computed to be
\begin{eqnarray}
  \mathrm{vol}_6 & = & \frac{1}{8\sqrt{3}} \gamma_1 \gamma_2 \gamma_3 \\
                 & = & (4\pi^2\alpha')^3
                       \sqrt{\frac{5^3|f_4^1 f_4^2 f_4^3|}{3^3|f_0|^3}} \,,
\end{eqnarray}
where the factor
\begin{equation} \label{StrangeFactors}
  \frac{1}{8\sqrt{3}}
  =       \biggl(\frac{1}{9^{1/6}}\biggr)^3 \biggl(\frac{\sqrt{3}}{9^{1/6}2}\biggr)^3  \, ,
\end{equation}
comes from the discrete identifications in the compact manifold. The four dimensional Planck length then becomes:
\begin{eqnarray}
  l_{P(4)} & = & \frac{1}{\sqrt{16\pi}} \left(\frac{\mathrm{vol}_6}{2\kappa_{10}^2 e^{2\varphi}} \right)^{-1/2} \\
           & = & |h_3| \sqrt{\frac{3^3 \alpha'}{2^7 5} \frac{|f_0|}{|f_4^1 f_4^2 f_4^3|}}\, ,
\end{eqnarray}
where we used the convention $2\kappa_{10}^2 = (2\pi)^7 \alpha'^4$.

%------------------------------------------------------------
\subsection{The orientifold in the DeWolfe {\it et al.} background}

The orientifold, constructed by modding out by $\Omega_p (-1)^{F_L}
\sigma$, lies on
\begin{equation}
  x_1 = x_3 = x_5 = 0\, .
\end{equation}
Taking into account the identifications under $\mathbb{Z}_3 \times
\mathbb{Z}_3$, we get the three dimensional surface along which the
orientifold is extended in the compact space. In figure
\ref{fig:O6-plane}, this surface is pictured in the fundamental
region of one of the tori of $T^2 \times T^2 \times T^2$. The
orientifold also fills the non-compact space.

The 3-cycle that is invariant under $\mathbb{Z}_2 \times \mathbb{Z}_3 \times \mathbb{Z}_3$,
is the cycle on which the orientifold is wrapped. This cycle $\alpha_0$, is determined by its Poincar\'e dual 3-form,
\begin{eqnarray} \label{OriginalOrientifold}
  \mathrm{O6} &:& \alpha_0\mathrm{-cycle}                             \\
              &:& \sqrt[4]{3} \sqrt{2} \,
                  \bigl(
                        dx_1 \wedge dx_3 \wedge dx_5
                    -   dx_2 \wedge dx_4 \wedge dx_5                  \nonumber \\
              & &   \qquad \;
                    -\, dx_1 \wedge dx_4 \wedge dx_6
                    -   dx_2 \wedge dx_3 \wedge dx_6
                  \bigr)                                              \\
              &=& \beta_0                                             \,.
\end{eqnarray}

\FIGURE{
  \centering

  \begin{picture}(0,0)
  \includegraphics{O6-6.pstex}
  \end{picture}

    \setlength{\unitlength}{3947sp}
    \begin{picture}(6040,4215)(2689,-5426)
    \put(5264,-3972){\makebox(0,0)[lb]{\smash{$\mathbb{Z}_3$}}}
    \put(6780,-3117){\makebox(0,0)[lb]{\smash{$\mathbb{Z}_3$}}}
    \put(6383,-1697){\makebox(0,0)[lb]{\smash{$\mathbb{Z}_2$}}}
    \put(4891,-5380){\makebox(0,0)[lb]{\smash{$\mathbb{Z}_2$}}}
    \put(8633,-4646){\makebox(0,0)[lb]{\smash{$x_1$}}}
    \put(3543,-1336){\makebox(0,0)[lb]{\smash{$x_2$}}}
    \end{picture}

  \caption{The $z_1$-plane of $T^6$ with the actions of the non-free $\mathbb{Z}_3$ and the
           orientifolding $\mathbb{Z}_2$ indicated. The O6-plane is pictured in thicker, dashed lines.}
  \label{fig:O6-plane}
}

%--------------------------------------------------------
\subsection{Remark on the $F_2$-flux}  \label{GregRemark}
The Bianchi identity for the massive type IIA solution reads:
\begin{equation}
  dF_2 = F_0 H_3 + 2 \kappa_{10}^2 \mu_p \delta_{\mathrm{O6}} \neq 0 \, ,
\end{equation}
where $\mu_p = -2 \sqrt{\pi} \kappa_{10}^{-1} (4\pi^2\alpha')^{-3/2}$, is the
charge of the orientifold. Equation (\ref{F2zero}),
\begin{equation}
  F_2 = 0
\end{equation}
should thus be seen as an approximation to the exact solution.

%---------------------------------------------------
\subsection{Scaling behavior} \label{ScalingSection}
The integer parameters $f_4^1, f_4^2$ and $f_4^3$ are not
constrained by any tadpole condition, but we need to take each $f_4^i \not= 0$, to
have a non-degenerate solution (see (\ref{gamma})).

We will be interested in the regime
\begin{equation}
  f_4^1 = f_4^2 = f_4^3 = N \, ,
\end{equation}
where we take $N \rightarrow \infty$. The parameters characterizing the compactification scale as:
\begin{eqnarray}
  l_{P(4)}         & \sim & N^{-\frac{3}{2}} \sqrt{\alpha'}  \\
  R_{\mathrm{AdS}} & \sim & N^{ \frac{9}{4}} l_{P(4)}        \\
  R_{\mathrm{KK}}  & \sim & N^{ \frac{7}{4}} l_{P(4)}                     \\
  g_s              & \sim & N^{-\frac{3}{4}}                 \, ,
\end{eqnarray}
where $R_{\mathrm{KK}} = \sqrt[6]{\mathrm{vol}_6}$ is a measure for
the size of the compact manifold.
We see that the string coupling $g_s$ is small, while the radii characterizing the
solution are large. In addition, we notice that the background remains effectively four dimensional since the AdS radius grows faster than the Kaluza Klein radius.

Let us now consider the regime where:
\begin{equation}
  f_4^1 = N, \qquad f_4^2 = f_4^3 = O(1) \, ,
\end{equation}
which results in,
\begin{eqnarray}
  l_{P(4)}         & \sim & N^{-\frac{1}{ 2}} \sqrt{\alpha'}  \\
  R_{\mathrm{AdS}} & \sim & N^{ \frac{3}{ 4}} l_{P(4)}        \\
  R_{\mathrm{KK}}  & \sim & N^{ \frac{7}{12}} l_{P(4)}        \\
  g_s              & \sim & N^{-\frac{1}{ 4}} \, .
\end{eqnarray}
However in this regime, $\gamma_1$ (see eq. (\ref{gamma})) shrinks to zero, indicating that (massive) type IIA is not the correct description for this case. The above scalings might thus not hold in this scaling limit.

%=====================================
%=====================================
\section{Approximate double T-duality}
The DeWolfe {\it et al.} model is formulated in massive type IIA SUGRA.
This theory does not have a perturbative world sheet expansion and
quantization of the $F_0$ is problematic. The second difficulty is
that the model also contains an orientifold which is a singular
object when described in type IIA SUGRA \cite{Schulz}. To study the
model, we will first apply two T-dualities, ignoring back reaction
of the orientifold. These will bring us to non-massive type IIA. We
will address the second problem by inserting the orientifold in the
dualized configuration. The T-duality transformations have the
additional benefit that the $H_3$ flux vanishes. The original $H_3$
flux turns into a geometric flux showing up as twists in the metric.

%---------------------------------------------------
\subsection{Approximate character of the T-dualities}
Let us first point out that applying a double T-duality on a
configuration with fluxes results in general in a non-geometric
compactification \cite{NonGeometric}. However, the T-dualities we
will perform are chosen such that we do not violate the condition
ensuring that we remain in the domain of geometric compactification
\cite{KSTT}.

We will first perform a T-duality on the $x_1$-direction, followed
by a T-duality in the $x_2$-direction. The T-duality transformations
will only be valid in an approximate sense:
\begin{itemize}
    \item The loops on the $T^6$ defined by the $x_1$ and $x_2$-directions are contractible
on the fixed points of the $\mathbb{Z}_2 \times \mathbb{Z}_3 \times
\mathbb{Z}_3$ identifications. We thus do not have an $S^1$-isometry
required for an exact T-duality.

    \item In addition, we will work in the approximation where $F_2 = 0$. As discussed
in section \ref{GregRemark}, this flux does not satisfy the Bianchi
condition. We thus expect that the Bianchi identity after
T-dualities will not be satisfied either. The $F_2$ flux is sourced
by the orientifold and by the flux term $F_0 H_3$. In the T-duality
computation we will keep track of the fluxes $F_0$, $H_3$ and the
cycle on which the orientifold is wrapped. This information will be
helpful to correct the T-dualized solution.

    \item Notice that we can expect the correct $F_2$ in the original
setup to depend on all coordinates $x_i$, since we expect that close
to the orientifold the $F_2$ flux resembles the $F_2$ flux of an
orientifold in flat space. Because of the identifications the
orientifold is extended along all coordinate directions (see figure
\ref{fig:O6-plane}). This implies that the $F_2$ flux breaks the
$S^1$-isometry in the coordinate directions. Thus, if we would
include the back reaction of the orientifold, which sources the
$F_2$-flux, we would not be able to T-dualize.

\end{itemize}

There are two related ways to view our approximate T-dualities. So
far we have emphasized the first, which is the interpretation of
T-duality mapping solutions of Type IIA supergravity into other
solutions.   In our case it takes solutions of massive Type IIA into
solutions of ordinary Type IIA, because the duality eliminates the
$F_0$ flux.   This duality is only approximate and in order to
perform it we must ignore the orientifold (or at least its back
reaction).

Alternatively, we can start from the orbifold conformal field theory
of DeWolfe {\it et al.} Turning on fluxes corresponds to
deformations of the background in the direction of certain vertex
operators, and the orientifold corresponds to modding out the CFT by
one of its symmetry operations.   We can perform an exact T-duality
on the CFT (just a change of variables) and try to understand which
vertex operators must be turned on in the T-dual
language\footnote{The real problem here is that since fluxes are
discrete, the sought for CFT is not a small perturbation of the
original orbifold.}. Similarly we can mod out by the T-dual symmetry
operation. The result of this operation is a CFT just as mysterious
as the one, one might have tried to write down in the original
picture.   However to leading order in the string tension expansion,
it leads to a new set of equations of motion, to which we may try to
find a solution.   The reader may choose whichever interpretation of
our procedure (s)he finds most convincing.   We do not pretend that we
have presented a rigorous argument for either approach.

At any rate, as a consequence of the approximations, we can expect
the solution after T-dualities to contain inconsistencies.  By
imposing the equations of motion, the Bianchi conditions and
supersymmetry conditions on the dualized configuration, we hope to
find a tractable version of the DeWolfe {\it et al.} solutions,
with a well defined large $N$ expansion.

\subsection{Orientifold projection in a twisted torus}
The $H_3$-flux in the DeWolfe {\it et al.} paper leads to twisting in the geometry after T-duality.
In this section we study how an orientifold fixed plane behaves under T-duality when a
non-trivial $H_3$-flux background is turned on.

Since the $H_3$-flux (\ref{DeWolfeSol1H3}) and the orientifold
(\ref{OriginalOrientifold}), have several components along different
$x_i$-directions, we get several twisting terms in the metric after
T-duality in the $x_i$ coordinate system. The T-duality action
allows us to break this problem in several smaller problems by
focusing on one term in the orientifold and one term of the
$H_3$-flux. Let us work out the case where we focus on the term
\begin{equation}
  H_3 \sim dx_1 \wedge dx_4 \wedge dx_6 + \ldots
\end{equation}
and where the orientifold fixed plane is wrapped on the compact 3-cycle
\begin{equation}
  \mathrm{O6} : dx_1 \wedge dx_3 \wedge dx_5 + \ldots
\end{equation}
in the Calabi-Yau manifold.
Let us rename and rescale, $x_1, x_4, x_6$ as $\theta, Y, Z$. We can now focus on
the 3-torus $T^3$ with $H_3$-flux \cite{KSTT}:
\begin{eqnarray}
  ds_3^2 & = & d\theta^2 + dY^2 + dZ^2 \\
  H_3    & = & d\theta \wedge dY \wedge dZ \, ,
\end{eqnarray}
where we take $\theta, Y$ and $Z$ to have periodicities $2\pi$. The location of the orientifold
in this $T^3$ subspace of the compact manifold is determined by the fixed plane of the symmetry,
\begin{eqnarray}
  \theta \, & \rightarrow & -\theta \\
  Y         & \rightarrow & Y       \\
  Z         & \rightarrow & Z       \, .
\end{eqnarray}

The bosonic part of the worldsheet action which encodes the dynamics on the $T^3$ is
given by ($d^2z = d\sigma^1 d\sigma^2$):
\begin{equation}
  S = \frac{1}{2\pi\alpha'} \int d^2z
        \{   \partial \theta \bar{\partial} \theta
           + \partial Y \bar{\partial} Y
           + \partial Z \bar{\partial} Z
           + Y \partial \theta \bar{\partial} Z
           - Y \partial Z \bar{\partial} \theta \} \, .
\end{equation}
This action is invariant under the periodic identifications of the
target space coordinates since the periodic identification, $Y
\rightarrow Y + 2\pi$, only contributes a total derivative. The
action is also invariant under the discrete orientifold symmetry
$\sigma \Omega_{\theta}$:
\begin{eqnarray}
  \theta(z,\bar{z}) & \rightarrow & -\theta(\bar{z},z) \\
  Y(z,\bar{z})      & \rightarrow &  Y(\bar{z},z)      \\
  Z(z,\bar{z})      & \rightarrow &  Z(\bar{z},z)      \, .
\end{eqnarray}

Now we can perform a T-duality in the $\theta$-direction by gauging the $U(1)$-isometry
along that direction \cite{WorldsheetTdualityReview}. The new action reads,
\begin{eqnarray}
  S & = & \frac{1}{2\pi\alpha'} \int d^2z
        \{   (\partial \theta + A)(\bar{\partial} \theta + \bar{A})
           + \partial Y \bar{\partial} Y
           + \partial Z \bar{\partial} Z \\
    &  & \qquad \qquad \quad
           + \,Y (\partial \theta + A) \bar{\partial} Z
           - Y \partial Z (\bar{\partial} \theta + \bar{A})
           + \tilde{\theta} F\} \, ,
\end{eqnarray}
with $F = \partial \bar{A} - \bar{\partial} A$. The field
$\tilde{\theta}$, is a Lagrange multiplier with period $2\pi$.
Integrating out $\tilde{\theta}$ gives the original action.

The new action is only invariant under the periodicity $Y
\rightarrow Y + 2\pi$, if we take $\tilde{\theta} \rightarrow
\tilde{\theta} - 2\pi Z$, simultaneously. We can also extend the
action of the orientifold symmetry by requiring that the new action
is invariant under the extended orientifold symmetry. The
orientifold symmetry, $\sigma \Omega_{\theta}$, becomes:
\begin{eqnarray}
  \theta(z,\bar{z})          \rightarrow  -\theta(\bar{z},z)         & \qquad &
  Y(z,\bar{z})               \rightarrow   Y(\bar{z},z)
\\
  A(z,\bar{z})               \rightarrow   -A(\bar{z},z)             & \qquad &
  Z(z,\bar{z})               \rightarrow   Z(\bar{z},z)
\\
  \bar{A}(z,\bar{z})         \rightarrow   -\bar{A}(\bar{z},z)       & \qquad &
  F(z,\bar{z})               \rightarrow   F(\bar{z},z)
\\
                                                                     & \qquad &
  \tilde{\theta}(z,\bar{z})  \rightarrow   \tilde{\theta}(\bar{z},z) \, .
\end{eqnarray}
Fixing the gauge with the condition $\theta = 0$ and integrating out
the fields $A$ and $\bar{A}$ gives the dual action:
\begin{eqnarray}
  S & = & \frac{1}{2\pi\alpha'} \int d^2z
        \{   (\partial \tilde{\theta} + Y\partial Z)
             (\bar{\partial} \tilde{\theta} + Y\bar{\partial} Z)
           + \partial Y \bar{\partial} Y
           + \partial Z \bar{\partial} Z \}  \, .
\end{eqnarray}

Translating this to the target space gives,
\begin{eqnarray}
  ds_3^2 & = & (d\tilde{\theta} + YdZ)^2 + dY^2 + dZ^2 \\
  H_3    & = & 0                                       \, ,
\end{eqnarray}
with the identifications,
\begin{eqnarray}
  \tilde{\theta} & \rightarrow & \tilde{\theta} + 2\pi  \\
  Y              & \rightarrow & Y + 2\pi \quad \mathrm{and}
      \quad \tilde{\theta} \rightarrow  \tilde{\theta} - 2\pi Z\\
  Z              & \rightarrow & Z + 2\pi  \, .
\end{eqnarray}
The metric is thus a circle bundle over a torus. The non-trivial
identifications indicate that the coordinate $\tilde{\theta}$, along
the fibre is not globally well-defined. Let us introduce the one
form $\Theta$, which is globally defined by $d\Theta = dY \wedge dZ$
and gives locally $\Theta = d\tilde{\theta} + YdZ$. The action of
the orientifold symmetry now reads:
\begin{eqnarray}
  \Theta & \rightarrow & \Theta  \\
  Y      & \rightarrow & Y       \\
  Z      & \rightarrow & Z       \, .
\end{eqnarray}
This is, for the example we considered, the orientifold wraps both
the fibre and the torus base space after T-duality. The Poincar\'e
dual form of the cycle on which the orientifold is wrapped becomes:
\begin{equation}
  \mathrm{O7} : dx_3 \wedge dx_5 + \ldots \, .
\end{equation}

We can repeat this exercise for different combinations of terms in
the $H_3$-flux and orientifold cycle. We find that the orientifold
either wraps the fibre of the twisted torus, $\Theta \rightarrow
\Theta$, or reflects the fibre $\Theta \rightarrow -\Theta$. In the
T-duality computation of the DeWolfe {\it et al.} solution, we will follow
the orientifold by keeping track of the Poincar\'e dual form of the
cycle on which the orientifold is wrapped. The above discussion
shows that this is a consistent treatment of the orientifold.

%--------------------------------------
\subsection{Doubly dualized background}
In appendix \ref{TdualDictionary}, we review the action of T-duality
on a background of SUGRA. In appendix \ref{TdualComputation}, we
work out the double T-duality transformation of the DeWolfe {\it et al.}
model. The result reads:
\begin{eqnarray}
ds^2     &=&   \frac{4\pi^2\alpha'}{\gamma_1}
               \left(9^{\frac{2}{3}} \, \Theta_1^2 + 4^2\,3^{-\frac{2}{3}} \, \Theta_2^2\right)
             + \gamma_2 (dx_3^2 + dx_4^2)
             + \gamma_3 (dx_5^2 + dx_6^2)                                                  \nonumber \\
         & & + \,ds_{\mathrm{AdS}_4}^2                                         \label{2xTdualSol1}   \\
H_3      &=& 0                                                                                       \\
e^{\varphi}
         &=& \frac{1}{4} |h_3| \sqrt[4]{\frac{3^3 5}{|f_0 f_4^1 f_4^2 f_4^3|}}                       \\
F_4      &=&    4 (2\pi\sqrt{\alpha'})^3 \sqrt[3]{3} f_4^1
                \frac{1}{\gamma_2\gamma_3} \mathrm{vol}_4                                            \\
F_2      &=& -  4 (2\pi\sqrt{\alpha'})^3 \frac{\sqrt[3]{3}}{\gamma_1}
                   (  f_4^2 \,dx_5 \wedge dx_6
                    + f_4^3 \,dx_3 \wedge dx_4 )                                           \nonumber \\
         & & + \, f_0 \frac{2\pi\sqrt{\alpha'}}{\gamma_1} \,
                4\cdot3^{\frac{1}{3}}\,\Theta_1 \wedge \Theta_2  \label{2xTDualF2}                   \\
F_0      &=&  0                                                                                      \\
\frac{1}{2\kappa_{10\tilde{A}}^2}
         &=& \frac{1}{2\kappa_{10A}^2} \frac{\gamma_1^2}{(4\pi^2\alpha')^2\,4\cdot3^{\frac{1}{3}}}                                 \\
\Theta_1 &=&   2\pi\sqrt{\alpha'} dx_1
             + 2\pi\sqrt{\alpha'} h_3
               \frac{\sqrt[4]{3}\sqrt{2}}{9^{\frac{1}{3}}}(x_3 dx_5 - x_4 dx_6)   \label{Thetas}    \\
\Theta_2 &=&   2\pi\sqrt{\alpha'} dx_2
             + 2\pi\sqrt{\alpha'} h_3
               \frac{\sqrt[4]{3}\sqrt{2}}{4\cdot3^{-\frac{1}{3}}}(-x_3 dx_6 - x_4 dx_5)  \label{2xTdualSol2}  \, ,
\end{eqnarray}
where $x_1 \in [0,9^{-1/6}]$ and $x_2 \in [0,2^{-1}\,3^{1/6}]$.

The original orientifold splits into an O5- and O7-plane after the
first T-duality (see appendix \ref{TdualComputation}). The second
T-duality recombines those two orientifold planes to give an
O6-plane wrapped on the Poincar\'e dual of the $\tilde{\alpha}_0$-cycle:
\begin{eqnarray}
  \mathrm{O6} &:    &  \sqrt[4]{3}\sqrt{2} \,
                        \Bigl(
                         + \frac{2\cdot3^{-\frac{1}{6}}\Theta_2}{2\pi\sqrt{\alpha'}\,9^{\frac{1}{6}}}
                           \wedge (dx_3 \wedge dx_5 - dx_4 \wedge dx_6)           \\
              &     &  \qquad \quad \; \,
                         + \frac{9^{\frac{1}{6}}\Theta_1}{2\pi\sqrt{\alpha'}\,2\cdot3^{-\frac{1}{6}}}
                           \wedge (dx_4 \wedge dx_5 + dx_3 \wedge dx_6)
                        \Bigr)                                                    \nonumber \\
              & =   &   \tilde{\beta}_0                                           \,.
\end{eqnarray}

%========================================================
%========================================================
\section{The Bianchi identity after the double T-duality}
As mentioned earlier, we expect the dualized solution
(\ref{2xTdualSol1})-(\ref{2xTdualSol2}) to contain inconsistencies.
We will do the full analysis of the consistency conditions later.
Here we will focus on the Bianchi condition. Taking the $F_2$ flux
from the dualized solution we compute:
\begin{eqnarray} \label{wrongF2}
  dF_2 & = &   f_0 \frac{2\pi\sqrt{\alpha'}}{\gamma_1} \,4\cdot3^{\frac{1}{3}}
                 (  d\Theta_1 \wedge  \Theta_2
                  -  \Theta_1 \wedge d\Theta_2)                                          \\
       & = &   f_0 h_3 \frac{(2\pi\sqrt{\alpha'})^3}{\gamma_1} \,2\cdot3^{\frac{1}{6}}
               \tilde{\beta}_0                                        \label{BetaTilde}  \, .
\end{eqnarray}
This 3-form is everywhere non-zero.

On the other hand, since the configuration after T-dualities is a
solution of massless type IIA string theory, with the orientifold as
only source for $F_2$, we expect the Bianchi identity to read:
\begin{equation} \label{2xTDualBianchi}
  \frac{1}{2\kappa_{10\tilde{A}}^2} dF_2
       = \mu_6 \delta_{\mathrm{O6}} = \mu_6 \delta(\tilde{\beta}_0)   \, .
\end{equation}
The distributional 3-form $dF_2$, is thus localized on the
orientifold plane, which lies on the Poincar\'e dual of the 3-form
$\tilde{\beta}_0$. This is clearly at odds with (\ref{BetaTilde}).
This inconsistency was not unexpected as mentioned earlier.

We will now modify the dualized background such that it satisfies
the Bianchi condition. From equation (\ref{2xTDualBianchi}) we get:
\begin{equation} \label{2xTDualBianchiBis}
  dF_2 = -2 \, \frac{(2\pi \sqrt{\alpha'})^3}{\gamma_1}
            \, 2 \cdot 3^{\frac{1}{6}} \, \delta(\tilde{\beta}_0) \, .
\end{equation}
Integration over the $\tilde{\beta}_0$-cycle gives,
\begin{equation}
  \int_{\tilde{\beta}_0} dF_2
  = \int_{\mathrm{vol}_6} dF_2 \wedge \ast_6 \tilde{\beta}_0
  = -2 \, \frac{(2\pi \sqrt{\alpha'})^3}{\gamma_1} \, 2 \cdot 3^{\frac{1}{6}}  \, ,
\end{equation}
or, after partial integration,
\begin{eqnarray}
          2\cdot3^{\frac{1}{6}}\, \frac{\gamma_1}{(2\pi\sqrt{\alpha'})^3}\, h_3
          \int_{\mathrm{vol}_6} F_2 \wedge (dx_3 \wedge dx_4 \wedge dx_5 \wedge dx_6)
    & = & -2                                                        \label{Int2xTDualBianchi}  \, .
\end{eqnarray}

In the original DeWolfe {\it et al.} solution, the flux from the
orientifold was absorbed by the $F_0 H_3$ term in the Bianchi
identity. This leads to the constraint $f_0 h_3 = -2$. The above
derivation shows how the twisted geometry, with the non-closed
3-form $\ast_6 \tilde{\beta}_0$, absorbs the orientifold flux without the
$F_0 H_3$ flux term.

The integrated Bianchi condition also gives us some information on the correct $F_2$
flux. It should contain a (distributional) term proportional to $f_0 \Theta_1 \wedge
\Theta_2$. We also learn that as $N \rightarrow \infty$, the $F_2$
flux has to decrease. From now on we will ignore the term,
\begin{equation}
  f_0 \frac{2\pi\sqrt{\alpha'}}{\gamma_1} \,4\cdot3^{\frac{1}{3}}\, \Theta_1 \wedge \Theta_2 \, ,
\end{equation}
in (\ref{2xTDualF2}). Instead we will include a term $F_{2, \,\mathrm{O6}}$ which satisfies (\ref{2xTDualBianchiBis}).

%=========================
%=========================
\section{Lift to M-theory}

%-----------------------------------------------------------------------------------------------------
\subsection{Entropy computation and motivation for an M-theory interpretation} \label{EntropyArgument}

Let us return to the original DeWolfe {\it et al.} solution. It is
supposed to be an $\mathrm{AdS}_4$ space-time, Maldacena dual to a
$2 + 1$ dimensional CFT. Following \cite{WittenBH} we can calculate
the entropy of this system,
\begin{eqnarray}
  S_{\mathrm{BH}} &   =  & \frac{A_{\mathrm{horizon}}^{\mathrm{Einst}}}{4 G_N^{\mathrm{4d}}}     \\
                  &   =  & \pi
                           \left(2 M R_{\mathrm{AdS}}\right)^{\frac{2}{3}}
                           \left(
                             \frac{160\cdot 2^{\frac{7}{8}} 3^{\frac{3}{4}} 5^{\frac{1}{4}} \pi} {27}
                             \frac{1}{|f_0|^{\frac{5}{4}} |h_3|^2}
                             |f_4^1 f_4^2 f_4^3|^{\frac{3}{4}}
                           \right)^{\frac{2}{3}}                                         \label{EntropyOriginal}     \, ,
\end{eqnarray}
where $M$ is the mass of the black hole.

Let us now consider the different scalings of the four form flux as discussed in section \ref{ScalingSection}.
For the regime $f_4^1 = f_4^2 = f_4^3 = N$, we find that
\begin{equation}
  S_{\mathrm{BH}} \sim  N^{\frac{3}{2}}  \, .
\end{equation}
This gives us the entropy as a function of the energy of the black
hole. Using standard CFT thermodynamics, this implies a scaling
\begin{equation}
  S_{\mathrm{BH}} \sim  N^{\frac{9}{2}}  \, .
\end{equation}
as a function of the temperature\footnote{This computation was done
independently in \cite{ahetal}}.  We do not know of a conformal
field theory with this kind of scaling.

If we take the other scaling regime where $f_4^1 \sim N$, while
the other two four form fluxes are held fixed, then we find an entropy scaling like $N^{3/2}$ at fixed
temperature, in the large $N$ limit. We know that the entropy of the CFT describing a stack
of $N$ M2-branes scales in precisely this manner\cite{EntropyM2}. This computation
thus seems to indicate that we should look for an M-theory
interpretation of the special cases of the DeWolfe {\it et al.}
background, with one large four form flux and the others of order
one. We note that repeating the entropy computation in the doubly
dualized background gives exactly the same answer as
(\ref{EntropyOriginal}).

Indeed, we found that in our double T-dual solution, there was four
form flux in the $\mathrm{AdS}_4$ directions, corresponding to of
order $N$ D2-branes at the end of the universe.  Note that this flux
comes only from $f_4^1$, the flux we have chosen to be large in
order to get M2-branes entropy scaling. Our entropy calculation
suggests that these D-branes are behaving like M2-branes. This could
be explained, if the IIA string coupling on the compact manifold is
strong in the region where the M2-branes are localized.

There is a second reason indicating that we should look for an
M-theory setting of the problem. The orientifold is a singular
object in 10D SUGRA. The M-theory lift of an orientifold in flat
space is the Atiyah-Hitchin manifold \cite{Seiberg} \cite{SW}. In 11
dimensions, we have thus a non-singular, completely geometric
description.  The orientifold is singular in the Type IIA limit,
because the string coupling is always large in the core of the
Atiyah-Hitchin manifold, no matter what its value is at infinity.
Since the orientifold locus includes the $\mathrm{AdS}_4$
directions, the D2-branes in our T-dual configuration are sitting
in a strong coupling region.  This explains why they behave like
M2-branes.

Such a picture is inconsistent with a weak coupling string theory
interpretation of the DeWolfe {\it et al.} configurations, with only one
large flux. We note that these observations are valid in the region
where $f_4^1$ is large, and the other two four form fluxes are
non-zero, and may be large or small. DeWolfe {\it et al.} only
claimed to have a weakly coupled four dimensional compactification
in the region where all fluxes are large. In our T-dual picture,
even this regime has a large number of branes sitting near the
orientifold.  Our next thought was that there might be an M-theory
interpretation with $N$ M2-branes embedded in a smooth manifold. We
will see that this is possible only for a single large flux with the
other fluxes fixed and non-zero.   Although the calculations of DeWolfe {\it et al.} still indicate a weakly coupled four dimensional
compactification in this limit, certain cycles of the compact
manifold shrink to zero for large $N$. These authors do not claim to
have control over the regime that we claim has a smooth M-theory
limit with comparable $\mathrm{AdS}_4$ and $M_7$ radii.

%----------------------------
\subsection{Lift to M-theory}
Given a massive type IIA solution (without $H_3$ flux), C.~M.~Hull
constructed a procedure to lift the solution to M-theory
\cite{MassiveHull}. This process consists roughly of a T-duality to
type IIB and then a lift via F-theory to M-theory\footnote{There are
complications on the quantum level with this construction
\cite{ModifiedHull}. Our analysis has been on the classical level.}.
As discussed earlier, if we T-dualize the DeWolfe {\it et al.} background
once, some $H_3$ flux remains which complicates the lift to
M-theory. Therefore, we will follow the slightly different track of
T-dualizing twice and then using the strong-weak correspondence
between type IIA string theory and M-theory to lift the
configuration to 11D\cite{Singh}.

In the 10D theories the orientifold is a singular object which we
included by keeping track of the cycle on which it was wrapped and
via its source term in the Bianchi identity. As mentioned earlier,
in M-theory this singular object translates into a non-singular
geometric object. Its explicit form is only known in the case of an
orientifold in flat space \cite{AHmanifold}. Our strategy will be  to first
construct a naive lift ignoring the orientifold. Appendix
\ref{IIA-MthDictionary} reviews the formulas to lift a non-singular
type IIA SUGRA background to M-theory. However, omitting the
orientifold will introduce inconsistencies. In a second step, we
will impose the consistency conditions and try to modify the naive
lift.

Constructing the naive lift to M-theory of the dualized background
gives, with $L_M = 2\pi \frac{\kappa_{10\tilde{A}}}{\sqrt{\pi}(2\pi\sqrt{\alpha'})^3}$:
\begin{eqnarray} \label{MtheoryLift}
ds^2     &=&   R_M^2 \Theta_M^2
              + \tilde{\gamma}_{11} \Theta_1^2
              + \tilde{\gamma}_{12} \Theta_2^2
              + \tilde{\gamma}_2 (dx_3^2 + dx_4^2)
              + \tilde{\gamma}_3 (dx_5^2 + dx_6^2)
              + ds_{\mathrm{AdS}_4}^2                                                       \\
G_4      &=&   6m\,\mathrm{vol}_4                                                           \\
\Theta_1 &=&   dx_1
             + h_3 \frac{\sqrt[4]{3}\sqrt{2}}{9^{\frac{1}{3}}} (x_3 dx_5 - x_4 dx_6)         \\
\Theta_2 &=&   dx_2
             + h_3 \frac{\sqrt[4]{3}\sqrt{2}}{4\cdot3^{-\frac{1}{3}}} (-x_3 dx_6 - x_4 dx_5)    \\
\Theta_M &=&   dx_M + A_1                                                                   \\
dA_1     &=&   - 2\cdot3^{\frac{1}{6}}
                 (f_4^2  dx_5 \wedge dx_6 + f_4^3  dx_3 \wedge dx_4)
               + \tilde{F}_{2, \,\mathrm{O6}}                         \label{dTheta_M}      \\
\frac{1}{2\kappa_{11M}^2}
         &=& \frac{1}{16\pi l_{P11}^9}
          =  \frac{1}{2\kappa_{10A}^2} \frac{\gamma_1^2}{(4\pi^2\alpha')^2\,4\cdot3^{\frac{1}{3}}}
             \frac{1}{L_M}  \, .
\end{eqnarray}
where $\tilde{F}_{2, \,\mathrm{O6}} = F_{2, \,\mathrm{O6}} / L_M$ and with
\begin{eqnarray}
R_M      &=& \frac{3^{\frac{5}{6}} \pi^{\frac{2}{9}}}{2^{\frac{7}{9}} 5^{\frac{1}{6}}}
             |f_0 h_3^4|^{\frac{1}{6}}
             \frac{|f_4^1|^{\frac{1}{6}}}{|f_4^2 f_4^3|^{\frac{1}{2}}}
             l_{P11}
                              \\
\tilde{\gamma}_{11}
         &=& \frac{2\cdot2^{\frac{4}{9}} 3^{\frac{5}{6}} \pi^{\frac{4}{9}} }{5^{\frac{1}{3}}}
             \frac{|f_0|^{\frac{1}{3}}}{|h_3|^{\frac{2}{3}}}
             |f_4^1|^{\frac{1}{3}}
             l_{P11}^2
                              \\
\tilde{\gamma}_{12}
         &=& \frac{32\cdot2^{\frac{4}{9}} 3^{\frac{5}{6}} \pi^{\frac{4}{9}} }{9\cdot5^{\frac{1}{3}}}
             \frac{|f_0|^{\frac{1}{3}}}{|h_3|^{\frac{2}{3}}}
             |f_4^1|^{\frac{1}{3}}
             l_{P11}^2
                              \\
\tilde{\gamma}_2
         &=& \frac{8\cdot 2^{\frac{4}{9}} 3^{\frac{5}{6}} 5^{\frac{2}{3}} \pi^{\frac{4}{9}}}{9}
             \frac{1}{|f_0 h_3|^{\frac{2}{3}}}
             |f_4^1|^{\frac{1}{3}} f_4^3
             l_{P11}^2
                              \\
\tilde{\gamma}_3
         &=& \frac{8\cdot 2^{\frac{4}{9}} 3^{\frac{5}{6}} 5^{\frac{2}{3}} \pi^{\frac{4}{9}}}{9}
             \frac{1}{|f_0 h_3|^{\frac{2}{3}}}
             |f_4^1|^{\frac{1}{3}} f_4^2
             l_{P11}^2
                              \\
R_{\mathrm{AdS}}
         &=& \frac{8\cdot 2^{\frac{2}{9}} 3^{\frac{5}{6}} 5^{\frac{5}{6}} \pi^{\frac{2}{9}}}{9}
             \frac{1}{|f_0|^{\frac{5}{6}} |h_3|^{\frac{4}{3}}}
             |f_4^1|^{\frac{1}{6}} |f_4^2 f_4^3|^{\frac{1}{2}}
             l_{P11}                                                                                                                   \\
m        &=& \frac{2^{\frac{7}{9}} 3^{\frac{7}{6}} 5^{\frac{1}{6}}}{160 \pi^{\frac{2}{9}}}
             |f_0|^{\frac{5}{6}} |h_3|^{\frac{4}{3}}
             \frac{f_4^1}{|f_4^1|^{\frac{7}{6}} |f_4^2 f_4^3|^{\frac{1}{2}}}
             \frac{1}{l_{P11}}                                                                      \label{MtheoryLiftm}               \, .
\end{eqnarray}
Notice that we also rescaled $\Theta_1, \Theta_2$ by $1/(2\pi\sqrt{\alpha'})$ as compared to the previous sections.

%==============================================
%==============================================
\section{Discussion of the naive M-theory lift}

%-----------------------------------------------
\subsection{Condition on $F_{2, \,\mathrm{O6}}$}
The term $L_M \tilde{F}_{2, \,\mathrm{O6}} = F_{2, \,\mathrm{O6}}$
in (\ref{dTheta_M}) has to satisfy the Bianchi identity
(\ref{2xTDualBianchiBis}). Integration as in
(\ref{Int2xTDualBianchi}) leads to the condition
\begin{equation} \label{F2ConditionMth}
    4\cdot3^{\frac{1}{3}} h_3 \int_{\mathrm{vol}_6}
                           \tilde{F}_{2, \, \mathrm{06}}
                           \wedge (dx_3 \wedge dx_4 \wedge dx_5 \wedge dx_6) = -2 \, .
\end{equation}
As discussed earlier, this constraint should lead to $f_0 h_3 = -2$.

%------------------------------------------
\subsection{Volume of the compact manifold}
The volume of the compact 7 dimensional manifold is
\begin{equation}
  \mathrm{vol}_7 = \frac{64\cdot 2^{\frac{5}{9}} 3^{\frac{5}{6}} 5^{\frac{5}{6}} \pi^{\frac{14}{9}}} {27}
                   \frac{1}{|f_0|^{\frac{5}{6}} |h_3|^{\frac{4}{3}}}
                   |f_4^1|^{\frac{7}{6}} |f_4^2 f_4^3|^{\frac{1}{2}}
                   l_{P11}^7 \, .
\end{equation}
The Kaluza-Klein radius becomes
\begin{eqnarray}
  R_{\mathrm{KK}} &=& \sqrt[7]{\mathrm{vol}_7}                                                                                          \\
                  &=& \frac{2^{\frac{59}{63}} 3^{\frac{29}{42}} 5^{\frac{5}{42}} \pi^{\frac{2}{9}}} {3}
                      \frac{1}{|f_0|^{\frac{5}{42}} |h_3|^{\frac{4}{21}}}
                      |f_4^1|^{\frac{1}{6}} |f_4^2 f_4^3|^{\frac{1}{14}}
                      l_{P11}                                                                                                           \, .
\end{eqnarray}

%-----------------------
\subsection{The entropy}
Let us express the entropy as a function of the energy of the black hole as in (\ref{EntropyOriginal}). The scaling part of the entropy of the configuration is given by,
\begin{eqnarray}
  S_{\mathrm{BH}}
     &\sim& \left(
               \frac{R_{\mathrm{AdS}}}{l_{P4}}
            \right)^{\frac{2}{3}}                                                                             \\
     &\sim& \left(
               \frac{160\cdot 2^{\frac{7}{8}} 3^{\frac{3}{4}} 5^{\frac{1}{4}} \pi} {27}
               \frac{1}{|f_0|^{\frac{5}{4}} |h_3|^2}
               |f_4^1 f_4^2 f_4^3|^{\frac{3}{4}}
            \right)^{\frac{2}{3}}                                                 \label{S_M}                 \, .
\end{eqnarray}
Comparing this to the scaling part of the entropy in the original
setup (see equation (\ref{EntropyOriginal})), we see that they match
perfectly.

%----------------------------
\subsection{Scaling behavior}
In the regime $f_4^1 = f_4^2 = f_4^3 = N$, the various parameters of the background scale as,
\begin{eqnarray}
  R_M              & \sim &  N^{-\frac{ 5}{ 6}} l_{P11}             \label{RmScalingAll} \\
  R_{\mathrm{AdS}} & \sim &  N^{ \frac{ 7}{ 6}} l_{P11}             \\
  \mathrm{vol}_7   & \sim &  N^{ \frac{13}{ 6}} l_{P11}^7           \\
  R_{\mathrm{KK}}  & \sim &  N^{ \frac{13}{42}} l_{P11}             \\
  m                & \sim &  N^{-\frac{ 7}{ 6}} l_{P11}^{-1}        \\
  S_{\mathrm{BH}}  & \sim &  N^{ \frac{ 3}{ 2}}                     \, ,
\end{eqnarray}
where we express the scaling of the entropy as a function of the energy as in (\ref{EntropyOriginal}).
We see that just as the original DeWolfe {\it et al.} solution, the
M-theory configuration is effectively 4 dimensional, since
$R_{\mathrm{AdS}}$ grows faster with $N$ than $R_{\mathrm{KK}}$.
Note that the same analysis on the doubly dualized type IIA background teaches us that background is also effectively 4 dimensional in this particular scaling regime.
The radii characterizing the solution grow as $N$ increases, except
the M-theory radius $R_M$, which decreases with growing $N$. We will discuss this property
below.

Taking a look at the other scaling $f_4^1 = N, f_4^2 = f_4^3 = O(1)$, we get,
\begin{eqnarray}
  R_M              & \sim &  N^{ \frac{ 1}{ 6}} l_{P11}             \label{RmScaling_1} \\
  R_{\mathrm{AdS}} & \sim &  N^{ \frac{ 1}{ 6}} l_{P11}             \\
  \mathrm{vol}_7   & \sim &  N^{ \frac{ 7}{ 6}} l_{P11}^7           \\
  R_{\mathrm{KK}}  & \sim &  N^{ \frac{ 1}{ 6}} l_{P11}             \\
  m                & \sim &  N^{-\frac{ 1}{ 6}} l_{P11}^{-1}        \\
  S_{\mathrm{BH}}  & \sim &  N^{ \frac{ 1}{ 2}}                     \, .
\end{eqnarray}
Here we conclude that the AdS and the compact manifold grow at the same rate such that
the compactification is not effectively four dimensional. On the other hand,
in this case all the radii of the 11 dimensional solution grow with $N$ making 11D SUGRA
a valid approximation for large $N$. As previously mentioned, the scaling of the entropy
as function of the temperature in this regime is $N^{3/2}$.

%-----------------------------------------------
\subsection{Checking the consistency conditions}
\subsubsection{M-theory equations of motion}
{}From (\ref{MtheoryAction}), we get the equation of motion for the metric:
\begin{equation}
  \mathrm{Ric}_{MN} = \frac{2}{4!}
                      \left(
                        G_{MPQR} \, G_{N}^{\ \ PQR} - \frac{1}{12} g_{MN} \, G_{PQRS} \, G^{PQRS}
                      \right)              \, .
\end{equation}
Taking the indices $M$, $N$ in the AdS space, this condition reduces to:
\begin{equation}
  \frac{1}{R_{\mathrm{AdS}}^2} = 4m^2        \label{NonCompactEom}  \, .
\end{equation}
This condition is satisfied as we can verify from
(\ref{MtheoryLiftm}).

The equation of motion for the compact part of the metric, $g^{(7)}_{mn}$, becomes:
\begin{equation} \label{eomCompactMetric}
  \mathrm{Ric}_{mn} = 6 m^2 g^{(7)}_{mn} \, .
\end{equation}

This implies that the compact 7-manifold is an Einstein manifold. As
is common in similar cases, this condition will be satisfied if the
supersymmetry condition is satisfied.

We can verify that the equation of motion and Bianchi condition on $G_4$,
\begin{eqnarray}
  d \ast G_4 + \frac{1}{2} G_4 \wedge G_4  &=&  0 \\
  d G_4                                    &=&  0 \, ,
\end{eqnarray}
are satisfied.

\subsubsection{Supersymmetry conditions}
The original background was an $\mathcal{N} = 1$ compactification in
four dimensions. From \cite{G2}, we learn that the supersymmetry
requirement on the M-theory lift, $\mathrm{AdS}_4 \times M_7$, is
that $M_7$ has weak $G_2$ holonomy. Weak $G_2$ holonomy of a
7-manifold is defined by the condition that there exists a 3-form
$\phi_3$ and a real number $m$ such that,
\begin{equation} \label{WeakG2}
  d\phi_3 = 4 m \ast_7 \phi_3 \, .
\end{equation}
{}From this condition one can derive the equation of motion (\ref{eomCompactMetric}) \cite{G2condition}.
We conclude that if the supersymmetry conditions are obeyed then the naive M-theory lift is fully consistent.

When $N \rightarrow \infty$, $m \rightarrow 0$ and the supersymmetry
condition on the compact manifold simplifies to $G_2$ holonomy:
\begin{eqnarray}
  d\phi_3         &=& 0 \\
  d\ast_7 \phi_3  &=& 0 \, .
\end{eqnarray}

The 4 dimensional analysis in \cite{DeWolfe} led to stricter
conditions on the signs of the $F_4$ flux parameters $f_4^1$,
$f_4^2$ and $f_4^3$ \footnote{Equation (\ref{KahlerCone}) follows
from the K\"ahler cone conditions for the background. There are
additional K\"ahler cone conditions for the blow ups of the
singularities. We do not consider those conditions here since our
strategy was to ignore the singularities in the first step.}:
\begin{eqnarray}
  \mathrm{sign}(f_0 f_4^1 f_4^2 f_4^3)        & < & 0                                         \\
  \mathrm{sign}(f_4^1) = \mathrm{sign}(f_4^2) & = & \mathrm{sign}(f_4^3)   \label{KahlerCone} \, .
\end{eqnarray}
Backgrounds violating the above condition are believed to be stable
but non-super-symmetric solutions \cite{DeWolfe}. We can thus expect
that the above conditions will follow from the weak $G_2$ holonomy
condition.

If we check the weak $G_2$ holonomy condition for the naive lift, we
find that it does not satisfy the conditions. We included the
implicitly determined flux $\tilde{F}_{2, \,\mathrm{O6}}$ which is sourced
by the orientifold, while we did not include the Atiyah-Hitchin like
geometry from the orientifold. As the coupling constant flows from
type IIA to M-theory, we expect the singular orientifold to get some
thickness, modifying the geometry in the region close to the
orientifold. We thus expect that the naive lift is only an
approximation for the geometry far away from the orientifold.

We did not succeed in finding an explicit solution to (\ref{WeakG2})
in the regime where $f_4^1 = N, f_4^2 = f_4^3 = O(1)$, but neither have we
found any obstruction to the existence of a metric of weak $G_2$
holonomy with the scaling properties and behavior near the M2-brane
locus that we were led to.  We believe that in the limit of a single
large flux there is a systematic M-theory expansion.  The background
is $AdS_4 \times M_7$, with $M_7$ a manifold of weak $G_2$ holonomy.
The anti de Sitter and $M_7$ radii are comparable.   For other
configurations of large flux we believe that the M-theory picture is
only valid locally, in the vicinity of the orientifold, but that
this region is large and cannot be ignored for large $N$.   The
string coupling {\it does} go to zero over another large region of
the manifold.

%--------------------------------------------------
\subsection{Interpretation as a stack of M2-branes}

The entropy argument of section \ref{EntropyArgument} indicated that
for a certain flux configuration we could expect the DeWolfe {\it et al.}
solution to be the near horizon of a stack of M2-branes. The
$\mathrm{AdS}_4 \times M_7$ background with a weak $G_2$ holonomy
condition on $M_7$, as discussed in the previous section, is in
\cite{G2} indeed interpreted as the near horizon limit of M2-branes.

The background of a stack of $N$ M2-branes at the tip of a cone, is given by
\begin{eqnarray}
  ds^2  &=& H^{-\frac{2}{3}} ds^2_{3} + H^{\frac{1}{3}} ds^2_{8}   \\
  G_4   &=& \mathrm{vol}_3 \wedge dH^{-1} \, ,
\end{eqnarray}
with $ds^2_{3}$ and $\mathrm{vol}_3$ the Minkowski metric and
worldvolume of the M2-branes and $ds^2_8 = du^2 + u^2 ds^2_7$ the
metric of the cone in the directions transverse to the M2-branes.
The function $H$ is given by
\begin{equation}
  H = 1 + \frac{a^6}{u^6} \, ,
\end{equation}
and $a$ is determined by the number of M2-branes \cite{NumberOfM2s}:
\begin{equation} \label{NumberOfM2sFormula}
  a^6 = N \frac{\kappa_{11M}^2 T_3}{3 \Omega_7}
      = N \frac{\kappa_{11M}^2}{3 a^{-7} \mathrm{vol}_7}
          \left(\frac{4\pi^2}{2\kappa_{11M}^2} \right)^{\frac{1}{3}} \, ,
\end{equation}
with $\mathrm{vol}_7$ the volume form on $ds^2_7$.
The near horizon limit of this background becomes (after a coordinate transformation $r = 2u^2/a$):
\begin{eqnarray}
  ds^2  &=& \frac{r^2}{4 a^2} (-dt^2 + dy_1^2 + dy_2^2) + \frac{a^2}{4r^2} dr^2 + a^2 ds^2_7 \\
  G_4   &=& \frac{6}{a} \mathrm{vol}_4 \, ,
\end{eqnarray}
where $\mathrm{vol}_4$ is the volume form on the $\mathrm{AdS}_4$
space which has $R_{\mathrm{AdS}} = a/2$. Comparing this to the
M-theory lift of the DeWolfe {\it et al.} solution (\ref{MtheoryLift}), we
find that $m = 1/a$ and using (\ref{NumberOfM2sFormula}), we compute
that the number of M2-branes is given by:
\begin{equation}
  N = |f_4^1| \, .
\end{equation}
We know that the entropy (as a function of energy) of the CFT
corresponding to a stack of M2-branes scales as $N^{1/2} =
|f_4^1|^{1/2}$. We can compare this to the scaling of the entropy of
the M-theory lift (\ref{S_M}), $|f_4^1 f_4^2 f_4^3|^{\frac{1}{2}}$.
In the regime where $f_4^1 = N$, $f_4^2 = f_4^3 = O(1)$, the scaling
of the entropy is exactly the same. We can thus interpret the
M-theory lift of the DeWolfe {\it et al.} solution in that regime
as a stack of $N$ M2-branes at the tip of a cone.  We do not have a
deep understanding of the more generic regime, where all four form
fluxes are large, nor the regime where two are large and one is
small.  We note again that De Wolfe {\it et al.} only claimed to
have a systematic weak coupling and low energy expansion when all
fluxes are large.

%-------------------------------------------------------------------------
\subsection{Validity of 11 dimensional supergravity in the generic regime}
Let us consider the regime $f_4^1 = f_4^2 = f_4^3 = N$. The M-theory radius of the naive lift
(\ref{RmScalingAll}) decreases as $N$ grows, indicating that the 11 dimensional
supergravity approximation cannot be trusted, since the curvature of
the background becomes too large and corrections to 11D SUGRA will
be important. We find that $N < 3$ for $R_M > l_{P11}$. However, we
also see that we need $N > 1$ for $\sqrt{\tilde{\gamma}_{11}} >
l_{P11}$. We see that supergravity is only valid in a certain small
range of values for $N$.

The above reasoning is entirely based on our naive lift. Including the orientifold in the geometry
changes the situation close to the orientifold. For an orientifold
embedded in flat space, the dilaton increases the closer you get to
the orientifold. This corresponds to a larger M-theory radius $R_M$.
We can expect the same behavior in our configuration: including the
correct geometry coming from the orientifold will give an M-theory
radius which is larger than our naive estimate.

The geometry of the 11D SUGRA solution that incorporates the
orientifold, can be thought of as an interpolation between the
region close to the orientifold (bolt-geometry)\cite{AHmanifold} and
the region far away from the orientifold (naive lift). The twisted
tori in the region away from the orientifold come from the $F_0 H_3$
term in the original Bianchi identity, while (part of) the twist in
the M-theory direction corresponds to the $F_2$ flux sourced by the
orientifold.

Let us now focus on large $N$. The size of the compact manifold is large,
so there are points located far away from the orientifold. We expect
this region far away from the orientifold to resemble our naive
lift. Further yet from the orientifold we enter into a weak coupling
region. This is the region where the original argument of DeWolfe
{\it et al.} operates. In the large flux limit there is a large
region where it fails.  To see this, note that our T-dual Type IIA
configuration has a flux in the $\mathrm{AdS}_4$ directions, consistent with
$N$ D2-branes lying in the orientifold locus.   The dilaton in such
a D2-brane background has the form
\begin{equation}
  e^{\varphi} = \left(1 + \frac{c_2 g_s N l_s^5}{r^5}  \right)^{1/4} \,,
\end{equation}
with $c_2$ a numerical constant.
Plugging in the DeWolfe {\it et al.} value for the coupling at
infinity, we find that the coupling gets large at a distance of
order $N^{1/20} l_s$ from the stack of D2-branes.   Thus, the weak
coupling approximation breaks down over a parametrically large
region of the manifold as $N \rightarrow \infty$.

\FIGURE{
  \centering

  \begin{picture}(0,0)
  \includegraphics{Mushroom5.pstex}
  \end{picture}

    \setlength{\unitlength}{3947sp}

\begin{picture}(5009,5815)(2101,-6777)
\put(2101,-1078){\makebox(0,0)[lb]{\smash{a.}}}
\put(2101,-3156){\makebox(0,0)[lb]{\smash{b.}}}
\put(2772,-4797){\makebox(0,0)[lb]{\smash{$N^{1/20} l_s$}}}
\put(3028,-5800){\makebox(0,0)[lb]{\smash{$R_{\mathrm{KK}}^{\mathrm{11d}}$}}}
\put(5455,-5021){\makebox(0,0)[lb]{\smash{$R_{\mathrm{KK}}^{\mathrm{10d}}$}}}
\put(4794,-2098){\makebox(0,0)[lb]{\smash{$R_{\mathrm{KK}}^{\mathrm{10d}}$}}}
\put(2880,-1494){\makebox(0,0)[lb]{\smash{$N^{1/20} l_s$}}}
\end{picture}%

  \caption{Two possible scenarios. The right hand side of each drawing represents the region where (massive) type IIA is the correct description and where the radius of curvature of the compact manifold, $R_{\mathrm{KK}}^{\mathrm{10d}}$, scales with $N$ at a slower rate than $R_{\mathrm{AdS}}$. The left hand side is the region close to the D2/orientifold locus (indicated with the dot). This region has a size that scales as $N^{1/20} l_s$. The compact manifold radii in this region, $R_{\mathrm{KK}}^{\mathrm{11d}}$, scale as fast as $R_{\mathrm{AdS}}$ scales with $N$, resulting in a flat patch (case a) or a mushroom cap (case b).}
  \label{fig:Mushroom}
}

If we make the plausible assumption that an 11D description is valid
near the orientifold, we find that the local radius of curvature in
the presence of the branes is of order the AdS radius.  We connect the
weak coupling geometry to a seven dimensional patch whose size is of
order $N^{1/20} l_s \sim N^{13/60} l_{P11}$ (see figure \ref{fig:Mushroom}), and whose geometry is
that of the manifold of weak $G_2$ holonomy we described above.  The
radius of curvature of that geometry is of order the AdS radius,
and thus, much larger than the size of the patch. There now
seem to be two possibilities for a geometrical description of what
is going on in the large $N$ limit\footnote{We emphasize that since we have {\it no} complete
approximation scheme for this regime, there is no argument that {\it
any} geometrical picture is valid.}. In the first, the patch is essentially flat (see figure \ref{fig:Mushroom}a).   Alternatively, the whole geometry could mushroom out to a large seven dimensional patch with
weak $G_2$ holonomy and size of order the AdS radius (see figure \ref{fig:Mushroom}b).   We believe that neither the methods of DeWolfe {\it et al.} nor our own, are
powerful enough to distinguish between these two alternatives. In
the second alternative there would be KK excitations with a mass of
order the inverse AdS radius, and the compactification would not be
four dimensional.

%====================================
%====================================
\section{Conclusions and speculation}
We believe that we have provided ground for suspecting that the
massive IIA description of the DeWolfe {\it et al.} background does not
provide a systematic low energy expansion due to the back reaction
of the orientifold. The doubly dualized description still has the
same problem. However, this low energy effective description has the
advantage that the $F_0$ and $H_3$ fluxes are absent. 

\paragraph{Regime $f_4^1 = N$, $f_4^2 = f_4^3 = O(1)$: \\}
The scaling of the entropy indicates that there might be a correct expansion using
11D SUGRA in this regime. We constructed a naive lift to 11 dimensions.  We gave
arguments that a 7-manifold of weak $G_2$ holonomy exists and that
$N$ M2-branes at an approximately Atiyah-Hitchin locus on this
manifold might give a description of the physics of these
compactifications. We reiterate that this is not a regime where DeWolfe {\it et al.} claimed to have a controlled expansion.

We thus claim that in the regime where $f_4^1$ is large, and the
other four form fluxes are of order $1$, there should be a valid 11D
SUGRA approximation to the DeWolfe {\it et al.} models.   This would be
the near horizon limit of the configuration of $f_4^1$ M2-branes at
the tip of a cone over a seven manifold $M_7$ of weak $G_2$
holonomy.  The linear size of $M_7$ scales in the same way as the
$\mathrm{AdS}_4$ radius. The exact description of this regime would
be a $2+1$ dimensional CFT with fixed temperature entropy of order
$(f_4^1)^{3/2}$.   It should be possible to find it as the endpoint
of the RG flow along a relevant perturbation of the CFT of M2-branes
in flat space, which breaks the symmetry down to minimal $2+1$
dimensional SUSY.   The supergravity solution would enable one to
compute dimensions of low dimension operators at this fixed point.
However, since the SUSY algebra is so small, there might not be any
checks of these computations at the UV fixed point.

There is no sense in which this model is well approximated by weakly
coupled string theory.  In addition, the compactification is not
approximately four dimensional.  The AdS and $M_7$ radii are
comparable.   If our picture is the correct one, the failure of the
weak coupling analysis should be attributed to the naive treatment
of the orientifold.   In M-theory, the center of the orientifold is a
locus of strong IIA coupling.   In these compactifications, for
large $f_1^4$ (in the T-duality frame we have chosen), a large
number of M2 branes sit at this locus, and their back reaction
completely changes the weak coupling geometrical picture.  Of
course, the limit of a single large flux was not controllable in the
picture of DeWolfe {\it et al.}.   Nonetheless it is striking that
a single shrinking cycle (from their point of view) can actually
lead to a completely different picture of the geometry, and of the
strength of the coupling.

\paragraph{Generic regime $f_4^1 = f_4^2 = f_4^3 = N$: \\}
This regime is more mysterious.  The fixed temperature entropy of the CFT scales like
$N^{9/2}$.  We would like to propose a heuristic explanation of this
scaling law, but we warn the reader that many aspects of this
proposal are obscure.   Klebanov and Tseytlin proposed an
explanation \cite{kleb} of the $N^3$ scaling of the $(2,0)$ CFT that
describes M5-branes, in terms of partially BPS states of membranes
in a pair of pants configuration with boundaries on three different
5-branes.   We would like to propose a similar explanation for the
generic scaling of the entropy in the models of DeWolfe et
al.   There are two important differences.  First of all, we
hypothesize multi-layered pairs of pants (see figure \ref{fig:KKpants}
for an illustration).  That is, each geometrical
pair of pants is wrapped by $N$ M2-branes rather than a single
one.   Secondly, the M2-branes end on Kaluza-Klein monopoles instead of on M5-branes. We claim that the
entropy comes from $N^3$ copies of the M2-brane field theory, each
with entropy $N^{3/2}$.

\FIGURE[t]{
  \centering

  \includegraphics[scale=0.70]{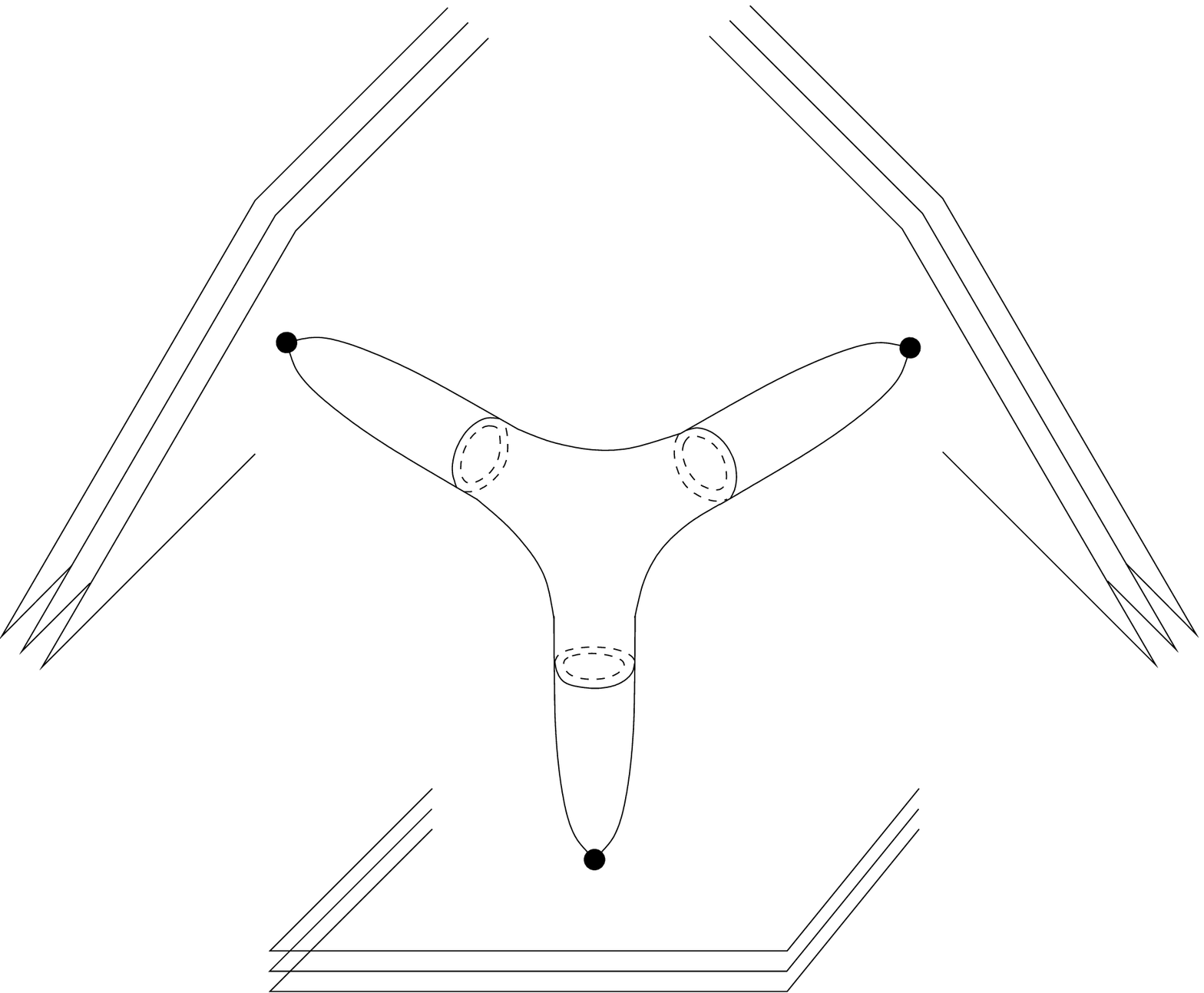}

  \caption{$N$ M2-branes ending on 3 stacks of $N$ Kaluza-Klein monopoles. Notice that each end of the trousers is stitched together.}
  \label{fig:KKpants}
}

The symmetry of the formulae under interchange of the three four form
fluxes, suggests that a picture based on string networks in Type IIB
string theory might capture some of the physics.   Thus, we would
imagine an Argentine bola string junction, with $N$ allowed sites
for each of its ends.   Each bola would consist of $N$ strings.  The
M2-brane scaling of the world volume theory of the string junction,
could be explained by hypothesizing that it passed through a large
volume on the compact manifold, where the M-theory torus had large
area.  The endpoints of the bola would be better described in terms
of weakly coupled Type IIB string theory, though perhaps different
ends would be weakly coupled in different S-duality frames.

We have argued that the weak coupling expansion claimed by DeWolfe
{\it et al.} in this regime cannot be uniformly valid, since there
is a region of size $N^{1/20} l_s$ where the coupling is not weak in
the large $N$ limit.  Since there is no single low energy field 
theory that describes these configurations, and since observables
in theories of gravity in Anti-de Sitter space are not local on the
compact manifold, we are not sure how one would go about making a 
systematic computation of these observables for large $N$.
% replaced the sentence below by the sentence above
%
%Of course, in string theory we do not really
%know how to construct observables which sample only local regions on
%a compact manifold. 
  
We presented two heuristic geometrical pictures of how the weak 
coupling geometry could connect on to a
region best described in terms of 11D SUGRA on a manifold of weak
$G_2$ holonomy.   The approximate 4 dimensionality of the
compactification is valid only in one of them. The weak coupling 
analysis {\it might} be missing a large {\it mushroom cap} region 
hidden near the strong coupling orientifold singularity.  We believe 
that no extant methods can distinguish between these two pictures, or
provide a systematic description of the physics of this system at
large $N$.   We have also presented a heuristic model of the entropy
of the regime with all fluxes large.   This model also depends on
the existence of large regions of the compact geometry which are
weakly curved eleven manifolds.

Given these arguments, and the success of our 11D picture in the
regime of a single large flux where the weakly coupled region completely disappears, we see serious reasons to doubt the validity of the simple weak coupling picture advocated by DeWolfe
{\it et al.}, even when all fluxes are large.  The reason for this
is the back reaction of a large number of branes near the strongly 
coupled orientifold locus, which changes the geometry in ways that 
cannot be understood from the perturbative picture.  

%  merged the paragraph below with a sentence above
%
%At a lower level of credibility, we have presented arguments that the compactification
%might not be approximately four dimensional when all fluxes are
%large.   
%The weak coupling analysis {\it might} be missing a large
%{\it mushroom cap} region hidden near the strong coupling
%orientifold singularity.

In our opinion, the best one could hope for would be some analog of
F-theory, in which different string expansions governed local
physics in different regions of the compact manifold. It is entirely
unclear to us whether the particular duality frame we have
emphasized is the best description of this regime. Furthermore,
since we are working with a very small amount of supersymmetry, it
is unlikely that we can use non-renormalization theorems to glean
exact information about these compactifications from their
geometrical formulation. This is a pity, because it is the only one
in which an approximately 4 dimensional compactification {\it might}
arise.

It would seem that the only way to really investigate the physics of
these backgrounds of string theory is to find and solve the dual $2
+ 1$ dimensional conformal field theory. For the special flux
configurations described above, it is plausible that this CFT can be
found by perturbing the Yang-Mills theory of D2-branes by an
appropriate relevant operator, obtaining the CFT dual to M2-branes
at the tip of a cone over a manifold of weak $G_2$ holonomy.  We
conjectured that the correct description of more general
configurations might be explained in terms of a tensor product of
field theories, or some theory which approximately reduced to such a
tensor product for purposes of counting the large $N$ asymptotics of
the entropy.   

\section{Acknowledgments}
One of the authors TB thanks M.~Dine and A.~Shomer for collaboration
on an initial attempt at this project. The authors also acknowledge
conversations with B.~Acharya, D.~Belov, F.~Denef, G.~Moore,
S.~Ramanujam, and W.~Taylor, which enhanced our understanding of
this difficult subject. We are particularly grateful to O. Aharony
for correcting an error in the original version of the manuscript.
KvdB also thanks A.~Garcia-Raboso, S.~Lukic and G.~Torroba for
useful discussions and is grateful for support and hospitality of
the SCIPP where part of this work was done. Finally, we would like
to thank the JHEP referee for pointing out that some of the
statements in the original version of the manuscript were incorrect.
This research was supported in part by DOE grant number
DE-FG03-92ER40689.

%%%%%%%%%%%%%%%%%%%%%%%%%%%%%%%%%%%%%%%%%%%%%%%%%%%%%%%%%%%%%%%%%%%%%%%%%%%%
%                      APPENDICES                                          %
%%%%%%%%%%%%%%%%%%%%%%%%%%%%%%%%%%%%%%%%%%%%%%%%%%%%%%%%%%%%%%%%%%%%%%%%%%%%
\appendix

%====================================================================================
%====================================================================================
\section{Appendix:  Type IIA - Type IIB T-duality dictionary} \label{TdualDictionary}
We take the bosonic type IIA action in the string frame to be (omitting the Chern-Simons terms):
\begin{eqnarray} \label{IIAaction}
  S_{\mathrm{IIA}} & = &   \frac{1}{2\kappa_{10A}^2}
                             \int \sqrt{-g_{10A}} e^{-2\varphi_A} (R + 4 \partial \varphi_A \partial \varphi_A)
                         - \frac{1}{4\kappa_{10A}^2}
                             \int e^{-2\varphi_A} H_3 \wedge \ast H_3                                                   \nonumber \\
                   &   & - \frac{1}{4\kappa_{10A}^2}
                             \int F_4 \wedge \ast F_4 + F_2 \wedge \ast F_2 + F_0 \wedge \ast F_0
                \, ,
\end{eqnarray}
while the bosonic type IIB action is given by:
\begin{eqnarray}
  S_{\mathrm{IIB}} & = &   \frac{1}{2\kappa_{10B}^2}
                             \int \sqrt{-g_{10B}} e^{-2\varphi_B} (R + 4 \partial \varphi_B \partial \varphi_B)
                         - \frac{1}{4\kappa_{10B}^2}
                             \int e^{-2\varphi_B} H_3 \wedge \ast H_3                                                   \nonumber \\
                   &   & - \frac{1}{4\kappa_{10B}^2}
                             \int \frac{1}{2} F_5 \wedge \ast F_5 + F_3 \wedge \ast F_3
                                            + F_1 \wedge \ast F_1           \, ,
\end{eqnarray}
where we impose the self-duality of $F_5$ in the equations of motion by hand (see \cite{Dima} for a consistent treatment of self-dual field theories).
The T-duality dictionary between both theories for a warped metric reads \cite{Buscher},
\begin{eqnarray}
  \begin{array}{rclcrcl}
  ds_A^2 &=& L_A^2 e^{ 2\alpha \phi} \Theta_A^2 + e^{2\beta \phi} ds_9^2 & \leftrightarrow &
  ds_B^2 &=& L_A^{-2} e^{-2\alpha \phi} \Theta_B^2 + e^{2\beta \phi} ds_9^2
  \\
  & & & & & &
  \\
  H_3    &=& \tilde{H}_3 + \tilde{H}_2 \wedge \Theta_A & \leftrightarrow &
  H_3    &=& \tilde{H}_3 - \tilde{F}^{NS} \wedge \Theta_B
  \\
  & & & & & &
  \\
  \varphi_A &&                                 & \leftrightarrow &
  \varphi_B &=& \varphi_A - \alpha \phi \, ,
  \\
  & & & & & &
  \\
  F_4    &=& \tilde{F}_4 + \tilde{F}_3 \wedge L_A \Theta_A & \leftrightarrow &
  F_5    &=& e^{(\alpha+\beta)\phi} \ast_9 \tilde{F}_4 + \tilde{F}_4 \wedge L_A^{-1}\Theta_B
  \\
  & & & & & &
  \\
  F_2    &=& \tilde{F}_2 + \tilde{F}_1 \wedge L_A \Theta_A & \leftrightarrow &
  F_3    &=& \tilde{F}_3 + \tilde{F}_2 \wedge L_A^{-1}\Theta_B
  \\
  & & & & & &
  \\
  F_0    &=& \tilde{F}_0                       & \leftrightarrow &
  F_1    &=& \tilde{F}_1 + \tilde{F}_0 \wedge L_A^{-1}\Theta_B
  \\
  & & & & & &
  \\
  \frac{1}{2\kappa_{10A}^2} &&                     & \leftrightarrow &
  \frac{1}{2\kappa_{10B}^2} &=& \frac{L_A^2}{2\kappa_{10A}^2}  \, ,
  \\
  & & & & & &
  \\
  d\Theta_A &=& \tilde{F}^{NS} & \leftrightarrow &
  d\Theta_B &=& -\tilde{H}_2 \, ,
  \\
  & & & & & &
  \\
  \Theta_A &=&  2\pi\sqrt{\alpha'} dx + \tilde{A}^{NS} & \leftrightarrow &
  \Theta_B &=&  2\pi\sqrt{\alpha'} dx - \tilde{B}_1 \, ,
  \end{array}
\end{eqnarray}
where the last line is valid locally, with $\tilde{F}^{NS} =
d\tilde{A}^{NS}$, $\tilde{H}_2 = d\tilde{B}_1$ and $x \in [0,1]$
parametrizes the $U(1)$ isometry. The Hodge star in $F_5$ is with
respect to the $ds_9^2$ metric on the type IIB side of the
dictionary. In our notation, the forms on the right hand sides of
the equations never contain $\Theta_A$ or $\Theta_B$ explicitly.

%==============================================================================================
%==============================================================================================
\section{Appendix: Computing the double T-dual of the DeWolfe {\it et al.} background} \label{TdualComputation}

We start from the DeWolfe {\it et al.} solution
(\ref{DeWolfeSol1})-(\ref{DeWolfeSol2}). We use the dictionary from
Appendix \ref{TdualDictionary}. To apply a first T-duality in the
$x_1$-direction, $x_1 \in [0,1]$, we take,
\begin{eqnarray}
  L_A^2           &=& \frac{\gamma_1}{4\pi^2\alpha'} 9^{-\frac{1}{3}}     \\
  \Theta_A        &=& 2\pi\sqrt{\alpha'} \, 9^{\frac{1}{6}} \, dx_1       \\
  \Theta_B        &=& 9^{\frac{1}{6}} \, \Theta_1                         \\
  \alpha = \beta  &=& 0                                                   \, .
\end{eqnarray}
The factor $9^{1/6}$ comes from the discrete symmetries $\mathbb{Z}_3 \times \mathbb{Z}_3$ (see (\ref{StrangeFactors})).
This results in:
\begin{eqnarray}
ds^2     &=&   \frac{4\pi^2\alpha'}{\gamma_1} \,9^{\frac{2}{3}}\, \Theta_1^2
             + \gamma_1 dx_2^2
             + \gamma_2 (dx_3^2 + dx_4^2)
             + \gamma_3 (dx_5^2 + dx_6^2)
             + ds_{\mathrm{AdS}_4}^2                                                               \\
H_3      &=& 4\pi^2\alpha' h_3 \sqrt[4]{3} \sqrt{2} \,
              (dx_2 \wedge dx_3 \wedge dx_6 + dx_2 \wedge dx_4 \wedge dx_5)                        \\
e^{\varphi_B}
         &=& \frac{1}{4} |h_3| \sqrt[4]{\frac{3^3 5}{|f_0 f_4^1 f_4^2 f_4^3|}}                     \\
F_5      &=&  4 (2\pi\sqrt{\alpha'})^3 \sqrt[3]{3} f_4^1
              \Bigl(
                  \ast_{9} (dx_3 \wedge dx_4 \wedge dx_5 \wedge dx_6)                   \nonumber  \\
         & &  \qquad \qquad \qquad \quad \; \;
                           + \,\sqrt{\frac{4\pi^2\alpha'\,9^{\frac{2}{3}}}{\gamma_1}}
                            dx_3 \wedge dx_4 \wedge dx_5 \wedge dx_6 \wedge \Theta_1
              \Bigr)                                                                               \\
         &=&  \left(
                4 (2\pi\sqrt{\alpha'})^3 \sqrt[3]{3} f_4^1
                \frac{1}{\gamma_2\gamma_3} \mathrm{vol}_4
              \right)
              \sqrt{\frac{\gamma_1}{4\pi^2\alpha'\,4\cdot3^{-\frac{1}{3}}}} \wedge
              (2\pi\sqrt{\alpha'}\,2\cdot3^{-\frac{1}{6}} dx_2)                         \nonumber  \\
         & & + \ast_{\tilde{9}}
              \left(
                4 (2\pi\sqrt{\alpha'})^3 \sqrt[3]{3} f_4^1
                \frac{1}{\gamma_2\gamma_3} \mathrm{vol}_4
              \right)                                                                              \\
F_3      &=& -  4 (2\pi\sqrt{\alpha'})^2 \sqrt[3]{3} \sqrt{\frac{4\pi^2\alpha'}{\gamma_1}}
                   (  f_4^2 \,dx_5 \wedge dx_6 \wedge dx_2
                    + f_4^3 \,dx_2 \wedge dx_3 \wedge dx_4 )                                       \\
F_1      &=&  \frac{f_0}{\sqrt{\gamma_1}}\, 9^{\frac{1}{3}}\, \Theta_1                             \\
\frac{1}{2\kappa_{10B}^2}
         &=& \frac{1}{2\kappa_{10A}^2} \frac{\gamma_1}{4\pi^2\alpha' \,9^{\frac{1}{3}}}            \\
\Theta_1 &=&   2\pi\sqrt{\alpha'} dx_1
             + 2\pi \sqrt{\alpha'} h_3
               \frac{\sqrt[4]{3} \sqrt{2}}{9^{\frac{1}{3}}} (x_3 dx_5 - x_4 dx_6)                  \, .
\end{eqnarray}
The last line is again only valid locally. We use the notation where, $\ast_9$,
means the Hodge star with respect to the metric after T-duality without
the $x_1$-direction, while, $\ast_{\tilde{9}}$, means the Hodge
star with respect to the metric after T-duality without the
$x_2$-direction. $\mathrm{vol}_4$ is the volume form of $AdS_4$.

The first T-duality transformation splits the original orientifold
(\ref{OriginalOrientifold}) into an O5- and an O7-plane:
\begin{eqnarray}
  \mathrm{O5} &:    &  \frac{\sqrt[4]{3} \sqrt{2}}{2\pi\sqrt{\alpha'}\,9^{\frac{1}{6}}} \,
                         (+ dx_3 \wedge dx_5
                          - dx_4 \wedge dx_6)                                          \\
  \mathrm{O7} &:    &  \frac{\sqrt[4]{3} \sqrt{2}\,9^{\frac{1}{6}}}
                                        {2\pi\sqrt{\alpha'}\,2\cdot3^{-\frac{1}{6}}} \,
                         (- dx_4 \wedge dx_5
                          - dx_3 \wedge dx_6) \wedge (2\pi\sqrt{\alpha'}
                            \,2\cdot3^{-\frac{1}{6}}dx_2) \wedge \Theta_1
\end{eqnarray}

After this first T-duality the solution still has an (approximate)
$U(1)$-isometry in the $x_2$-direction, $x_2 \in [0,\sqrt{3}/2]$. We
take
\begin{eqnarray}
  L_A^2           &=& \frac{4\pi^2\alpha'}{\gamma_1}
                      \left(\frac{4}{3} \, 9^{\frac{1}{3}} \right)                    \\
  \Theta_B        &=& 2\pi\sqrt{\alpha'}
                      \left(\frac{2}{\sqrt{3}} \, 9^{\frac{1}{6}}\right) dx_2         \\
  \Theta_A        &=& 2\cdot3^{-\frac{1}{6}}\, \Theta_2                                  \\
  \alpha = \beta  &=& 0                                                               \, ,
\end{eqnarray}
this results in the solution (\ref{2xTdualSol1})-(\ref{2xTdualSol2}).

%============================================================================
%============================================================================
\section{Appendix: Type IIA - M-theory dictionary}  \label{IIA-MthDictionary}

For the type IIA theory we start again from the action
(\ref{IIAaction}), for M-theory we take as definition of our theory,
\begin{eqnarray} \label{MtheoryAction}
  S_{\mathrm{M}} & = &  \frac{1}{2\kappa_{11}^2} \int \sqrt{-g_{11}} R
                       -\frac{1}{4\kappa_{11}^2} \int F_4 \wedge \ast F_4                                \\
                 &   & - \frac{1}{4\kappa_{11}^2}
                         \int C_3 \wedge F_4 \wedge F_4
\, .
\end{eqnarray}

The compactification of M-theory on a circle gives the following
type IIA - M-theory correspondence:
\begin{eqnarray}
  \begin{array}{rclcrcl}
  ds_A^2 &=& ds_{10}^2                                                                                & \leftrightarrow &
  ds_M^2 &=& L_M^{2} e^{\frac{4}{3} \varphi_A} \Theta_M^2 + e^{-\frac{2}{3}\varphi_A} ds_{10}^2
  \\
  & & & & & &
  \\
  \varphi_A &&&&
  &&
  \\
  & & & & & &
  \\
  H_3    && & \leftrightarrow &
  G_4    &=& F_4 + H_3 \wedge L_M \Theta_M
  \\
  & & & & & &
  \\
  F_4       &&&&
  &&
  \\
  & & & & & &
  \\
  F_2       &&                                     & \leftrightarrow &
  d\Theta_M &=& \frac{1}{L_M} F_2   \, ,
  \\
  & & & & & &
  \\
  &&&&
  \Theta_M &=& dx_M + \frac{1}{L_M}C_1 \, ,
  \\
  & & & & & &
  \\
  F_0       &=& 0  &&
  &&
  \\
  & & & & & &
  \\
  \frac{1}{2\kappa_{10A}^2} & & & \leftrightarrow &
  \frac{1}{2\kappa_{11M}^2} &=& \frac{1}{2\kappa_{10A}^2 L_M}  \, ,
  \\
  & & & & & &
  \\
  & & & &
  L_M &=& 2\pi \frac{\kappa_{10A}}{\sqrt{\pi}(2\pi\sqrt{\alpha'})^3} \, ,
  \end{array}
\end{eqnarray}
with $x_M \in [0,1]$.

%%%%%%%%%%%%%%%%%%%%%%%%%%%%%%%%%%%%%%%%%%%%%%%%%%%%%%%%%%%%%%%%%%%%%%%%%%%%
%                      REFERENCES                                          %
%%%%%%%%%%%%%%%%%%%%%%%%%%%%%%%%%%%%%%%%%%%%%%%%%%%%%%%%%%%%%%%%%%%%%%%%%%%%

\end{document}